\documentclass[showpacs,showkeys,preprintnumbers,superscriptaddress,aps]{revtex4}
\usepackage{amsmath}
\usepackage{dcolumn}
\usepackage{bm}
\usepackage{graphicx}
\usepackage{color}

\definecolor{rot}{rgb}{0.75,0.05,0.25}
\definecolor{hellgrau}{gray}{0.5}
\definecolor{blau}{rgb}{0,0,0.7}

\begin{document}

\title{Maximum Entropy Principle for the Microcanonical Ensemble}
\author{Michele Campisi}
\email{campisi@unt.edu}
\author{Donald H. Kobe}
\email{kobe@unt.edu:}
\affiliation{Department of Physics,University of North Texas Denton, TX 76203-1427, U.S.A.}
\date{\today }

\begin{abstract}
We derive the microcanonical ensemble from the Maximum Entropy
Principle (MEP) using the phase space volume entropy of P. Hertz.
Maximizing this entropy with respect to the probability
distribution with the constraints of normalization and average
energy, we obtain the condition of constant energy. This approach
is complementary to the traditional derivation of the
microcanonical ensemble from the MEP using Shannon entropy and
assuming a priori that the energy is constant which results in
equal probabilities.
\end{abstract}

\pacs{05.30.Ch, 05.30.-d, 05.20.Gg, 89.70.+c}
\keywords{microcanonical ensemble, maximum entropy principle, constraints,
quantum ensemble, classical ensemble, probability distribution}
\maketitle

\section{Introduction}

The seminal works of Jaynes \cite{jaynes1, jaynes2} presents the information
theory approach to statistical physics using the Maximum Entropy Principle
(MEP). In the original papers, Jaynes maximized the Shannon information
entropy using constraints of normalization and average energy to obtain the
canonical ensemble. Later on, Tsallis \cite{tsallis1} maximized generalized
information entropies, like the R\'{e}nyi and Tsallis entropies, using
constraints of normalization and average energy to obtain deformed
exponential distributions that describe the behavior of nonextensive systems.

In this paper we show that there is also a special information entropy
associated with the microcanonical ensemble. This microcanonical information
entropy is the phase-space volume entropy, originally due to P. Hertz \cite%
{hertz} (see also \cite{gibbs}) that satisfies the heat theorem
\cite{boltz1,campisi1, gallavotti}). Using this entropy in the MEP
with constraints of normalization \emph{and} average energy, we
obtain the condition that the energy distribution is a delta
function, \emph{i.e.}, we derive the microcanonical ensemble from
the MEP.

In Section 2 we review the traditional application of the MEP to
the microcanonical ensemble. The quantum statistical application
of the MEP with discrete probabilities using the volume entropy is
treated in Section 3. The classical statistical application is
given in Section 4, which employs integration and functional
differentiation with continuous probability distribution
functions. The conclusion is given in Section 5.

\section{Traditional Approach to the Microcanonical Ensemble}

The traditional MEP is reviewed here to contrast it with our
approach and to establish the notation. The traditional approach
to the quantum microcanonical ensemble starts with the assumption
that the system is isolated and has a fixed energy $U$. Such a
macrostate of energy $U$ can be realized in a number $W$ of
possible ways each corresponding to a microstate $i$. Then one
looks for the probability $p_{i}$ that the system is in a certain
state $i$ with energy $U$. In quantum mechanics $U$ is an
eigenvalue of the Hamiltonian operator $E_{\beta }$ and $W$ is its
degeneracy $g_{\beta }$, \emph{i.e}., $U=E_{\beta }$ and
$W=g_{\beta }$. Since we are looking for the probability of a
state $i$ that is already assumed to belong to the eigenvalue
$E_{\beta }$, the traditional MEP does not have to use the energy
constraint and is

\begin{equation}
-\sum_{j\in \{j|E_{j}=E_{\beta }\}}p_{j}\log p_{j}-\lambda \left( \sum_{j\in
\{j|E_{j}=E_{\beta }\}}p_{j}-1\right) =\text{ maximum,}  \label{eq:fakeMEP}
\end{equation}%
where the first term is Shannon entropy and the sums are over states
restricted to $j\in \{j\left| E_{j}=E_{\beta }\}\right. .$

The MEP in (\ref{eq:fakeMEP}) gives Laplace's Principle of Insufficient
Reason
\begin{equation}
p_{i}=\frac{1}{g_{_{\beta }}}=\text{constant for \ }i\in \{j\left|
E_{j}=E_{\beta }\}\right. ,  \label{fmcp}
\end{equation}%
that shows the states $j$ in the given macrostate $\beta $ with energy $%
E_{\beta }$ are equiprobable. Thus the maximization procedure
gives us a \emph{flat} distribution. With some abuse of
terminology Eq. (\ref{fmcp}) is often referred to as the
``microcanonical ensemble,'' but it is defined only for the states
$j$ such that $E_{j}=E_{\beta }.$ Strictly speaking, the
microcanonical ensemble is defined on the whole phase space and
constrains the system state to lie on a given surface of constant
energy. The microcanonical ensemble of energy $E_{\beta }$ is
really given as \cite{Ruelle}%
\begin{equation}
p_{i}=\frac{1}{g_{\beta }}\delta _{Kr}(E_{i},E_{\beta })  \label{eq:microCan}
\end{equation}%
where $\delta _{Kr}$ is the Kronecker delta [$\delta _{Kr}(x,y)=1$
for $x=y$ and $0$ for $x\neq y$]. The Kronecker delta does not
appear in Eq. (\ref {fmcp}) because it is assumed \emph{a priori}.

We stress that the traditional approach does not maximize on the whole set
of eigenstates of the Hamiltonian but rather on the subset of eigenstates
belonging to the eigenvalue $E_{\beta }$. This approach is quite different
from Jaynes's derivation of the canonical ensemble, where $i$ runs over
\emph{all} the energy eigenstates. In the following section we ask the
question: \emph{Is it possible to derive the microcanonical ensemble in (\ref%
{eq:microCan}) from a suitable MEP performed on the whole set of
eigenstates, as Jaynes did for the canonical ensemble?}

\section{Derivation of the Microcanonical Distribution: Quantum Case}

In order to answer to the question posed above, let us proceed by analogy
with Jaynes's approach to the canonical ensemble. In order to obtain the
canonical distribution,
\begin{equation}
p_{i}=Z^{-1}e^{-\beta E_{i}}.  \label{eq:Can}
\end{equation}%
where $Z$ is the partition function, and $\beta ^{-1}$ is the absolute
temperature, one maximizes the Shannon entropy $-\sum_{i}p_{i}\log p_{i}$
under the energy constraint $U=\sum_{i}p_{i}E_{i}$ and the normalization
constraint $\sum_{i}p_{i}=1,$ where $i$ runs over all energy eigenstates.
When the Shannon entropy is evaluated with the maximal distribution (\ref%
{eq:Can}) we obtain the correct \emph{thermodynamic entropy}
\begin{equation}
\beta U+\log \sum_{n}e^{-\beta E_{n}}.  \label{eq:can-entropy}
\end{equation}%
This thermodynamic entropy is correct in the sense that it satisfies the
\emph{heat theorem} whenever the averages are calculated over the canonical
ensemble \cite{campisi-dual}.

In the microcanonical case the correct \emph{thermodynamic
entropy} that satisfies the \emph{heat theorem} is given by the
logarithm of the volume of phase space enclosed by the
hypersurface of
energy $U=E_{\beta }$ \cite{campisi1,campisi-dual}. In the quantum version such entropy is%
\begin{equation}
S(U)=\log \Phi (U)\doteq \log \sum_{j}\theta (U-E_{j}),  \label{Phi}
\end{equation}%
where $\theta (x)$ is the step function [$\theta (x)=1$ for $x\geq 0,$ and $0
$ for $x<0]$.

Since we are now performing the maximization on the totality of eigenstates,
we \emph{must} use the energy constraint as we do with the canonical
ensemble. Thus we are maximizing (\ref{Phi}) under the normalization \emph{%
and} average energy conditions,
\begin{equation}
\sum_{j}p_{j}=1,\quad \sum_{j}p_{j}E_{j}=U,  \label{constr}
\end{equation}%
Using the constraints in Eq. (\ref{constr}), we can rewrite the entropy in
Eq. (\ref{Phi}) as
\begin{equation}
S(p)=\log \sum_{j}\theta \left( \sum_{k}p_{k}E_{k}-E_{j}\sum_{k}p_{k}\right)
\label{Sconstr}
\end{equation}%
where the sums on $j$ and $k$ are over all states. The discrete probability
distribution $p=\{p_{i}\}$ for the microcanonical ensemble is obtained when
this entropy is an extremum. Differentiating Eq. (\ref{Sconstr}) with
respect to $p_{i}$ and setting the result equal to zero, we obtain
\begin{equation}
\frac{\partial S}{\partial p_{i}}=\frac{1}{\Phi (U)}\sum_{j}\delta \left(
U-E_{j}\right) (E_{i}-E_{j})=0,  \label{dS}
\end{equation}%
for each state $i$, where $\theta ^{\prime }(x)=\delta (x)$ is the Dirac
delta function. We can see by inspection that Eq. (\ref{dS}) is satisfied if
$E_{j}\neq U.$ When $E_{j}=U$ the state $i$ must be such that $E_{i}=E_{j}$ $%
[$because\emph{\ }$x\delta (x)=0].$ In the latter case we have $E_{i}=U$.
The probability distribution for states $i$ is therefore%
\begin{equation}
p_{i}=A_{i}\text{ }\delta _{Kr}(E_{i},U=E_{\beta }),  \label{prob}
\end{equation}%
where $A_{i}$ are yet to be determined. The Kronecker delta $\delta
_{Kr}(E_{i},E_{\beta })$ imposes the restriction that the probability of
states $i\notin \{i|E_{i}=E_{\beta }\}$ are zero.

Since there is nothing to distinguish different states $i\in
\{i|E_{i}=E_{\beta }\},$ we can invoke Laplace's Principle of Insufficient
Reason, obtained from the traditional MEP approach, to choose $%
A_{i}=A_{\beta }$ to be the same for all states belonging to the
same eigenenergy $E_{\beta }$. Using the constraint of
normalization in Eq. (\ref{constr}), we obtain
\begin{equation}
p_{i}=\frac{1}{g_{\beta }}\delta _{Kr}(E_{i},E_{\beta }),  \label{p}
\end{equation}%
which is the microcanonical probability distribution. The only nonzero
contributions are from states $i$ with fixed energy $E_{i}=E_{\beta }$.

\section{Derivation of the Microcanonical Distribution: Classical Case}

The derivation of the classical microcanonical distribution proceeds in a
way analogous to the quantum derivation. Because we need to use a continuous
probability distribution, we must use integration and functional
differentiation in the MEP. However, the treatment is sufficiently different
to merit some discussion.

Equation (\ref{dS}) for the classical volume entropy of P. Hertz \cite{hertz}
is
\begin{equation}
S(U)=\log \Phi (U),  \label{Scl}
\end{equation}%
where $U$ is again the energy. In the classical case the function $\Phi (U)$
is now the volume of phase space enclosed by the hypersurface of energy $U$ %
\cite{campisi1}
\begin{equation}
\Phi (U)\doteq \int_{\mathbf{z\in \{z|}H(\mathbf{z})\leq U\}}d\mathbf{z}%
=\int d\mathbf{z}\text{ }\theta (U-H(\mathbf{z})),  \label{Phicl}
\end{equation}%
where the Hamiltonian is $H(\mathbf{z})$ and the step function $\theta (U-H(%
\mathbf{z}))$ provides the limits for the integral. The phase space
coordinate $\mathbf{z}=(\mathbf{q,p})$ consists of the set of canonical
coordinates $\mathbf{q}=\{q_{i}\}_{i=1}^{n}$ in $n$-dimensional space and
the set of their conjugate canonical momenta $\mathbf{p}=\{p_{i}\}_{i=1}^{n}%
\mathbf{.}$ The element of volume in $2n$-dimensional phase space is $d%
\mathbf{z=}d^{n}q$ $d^{n}p$ and integration is over all phase space if no
limits are shown.

For the classical case, the constraints on normalization and average energy
corresponding to Eq. (\ref{constr}) are%
\begin{equation}
\int d\mathbf{z}\rho (\mathbf{z})=1,\text{ \ \ }\int d\mathbf{z}\rho (%
\mathbf{z})H(\mathbf{z})=U,  \label{Constr}
\end{equation}%
respectively, where $\rho (\mathbf{z})$ is the probability density in phase
space. The MEP for the classical microcanonical ensemble is analogous to the
quantum case. Using Eq. (\ref{Phicl}) and the constraints of normalization
and average energy in Eq. (\ref{Constr}), we can rewrite the entropy in Eq. (%
\ref{Scl}) as a functional%
\begin{equation}
S[\rho ]=\log \int d\mathbf{z}^{\prime }\text{ }\theta \left( \int d\mathbf{z%
}^{\prime \prime }\rho (\mathbf{z}^{\prime \prime })H(\mathbf{z}^{\prime
\prime })-H(\mathbf{z}^{\prime })\int d\mathbf{z}^{\prime \prime }\rho (%
\mathbf{z}^{\prime \prime })\right) ,  \label{SConstr}
\end{equation}%
where the integration is over all phase space. The continuous probability
distribution $\rho =\rho (\mathbf{z})$ for the microcanonical ensemble is
obtained when this entropy is an extremum. Functionally differentiating Eq. (%
\ref{SConstr}) with respect to $\rho (\mathbf{z})$ and setting the result
equal to zero, we obtain%
\begin{equation}
\frac{\delta S[\rho ]}{\delta \rho (\mathbf{z})}=\frac{1}{\Phi (U)}\int d%
\mathbf{z}^{\prime }\delta \left( U-H(\mathbf{z}^{\prime })\right) \left( H(%
\mathbf{z})-H(\mathbf{z}^{\prime })\right) =0.  \label{DelS}
\end{equation}%
By inspection we see that this equation is satisfied if $\mathbf{z}^{\prime }
$ is such that $H(\mathbf{z}^{\prime })\neq U$. When $H(\mathbf{z}^{\prime
})=U$ for some values $\mathbf{z}^{\prime }$we must also have $H(\mathbf{z}%
^{\prime })=H(\mathbf{z})$ for some values of $\mathbf{z}$ $[$because\emph{\
}$x\delta (x)=0].$ In the latter case we therefore have $H(\mathbf{z})=U.$
The distribution function $\rho (\mathbf{z})$ therefore has a delta function
that restricts the Hamiltonian to the hypersurface of energy $U$,
\begin{equation}
\rho (\mathbf{z})=A(\mathbf{z})\text{ }\delta \left( U-H(\mathbf{z})\right) ,
\label{rhoun}
\end{equation}%
where $A(\mathbf{z})$ is an arbitrary function of $\mathbf{z,}$ Since there
is nothing to distinguish different points in phase space $\mathbf{z}\in \{%
\mathbf{z}|H(\mathbf{z})=U\}$ that are all on the energy hypersurface,\ we
can invoke Laplace's Principle of Insufficient Reason to choose $A(\mathbf{z}%
)=A_{U},$ which is constant for fixed $U$ for all these phase space points.\
\ The normalization condition in Eq. (\ref{Constr}) then becomes%
\begin{equation}
\int d\mathbf{z}\rho (\mathbf{z})=\int d\mathbf{z}\text{ }A(\mathbf{z})\text{
}\delta \left( U-H(\mathbf{z})\right) =A_{U}\int d\mathbf{z}\text{ }\delta
\left( U-H(\mathbf{z})\right) =1.  \label{Normal}
\end{equation}%
The last integral in Eq. (\ref{Normal}) can be performed by making a change
of variables to $e=H(\mathbf{z}),$ which gives
\begin{equation}
\int d\mathbf{z}\text{ }\delta \left( U-H(\mathbf{z})\right) =\int de\frac{d%
\mathbf{z}}{de}\delta \left( U-e\right) =\left( \frac{d\mathbf{z}}{de}%
\right) _{e=U}\equiv \Omega (U),  \label{DofS}
\end{equation}%
where the function $\Omega (U)$ is the density of states for
energy $U$, \emph{i.e.}, the number of states per unit energy.
Substituting Eq. (\ref {DofS}) into Eq. (\ref{Normal}), we obtain
$A_{U}=\Omega (U)^{-1}.$ Therefore, the probability distribution
function in phase space in Eq. (\ref{rhoun}) becomes%
\begin{equation}
\rho (\mathbf{z})=\frac{1}{\Omega (U)}\delta \left( U-H(\mathbf{z})\right) ,
\label{rhocl}
\end{equation}%
which is in fact the well-known classical microcanonical distribution. If
the phase space point $\mathbf{z}$ is not on the energy hypersurface $U=H(%
\mathbf{z})$ the probability density is zero. This probability density is
analogous to the probability distribution in Eq. (\ref{p}) for the quantum
case, where the degeneracy $g_{\beta }$ corresponds to the density of states
$\Omega (U)$.

\section{Conclusion}

In this work we have reviewed the traditional
information-theoretic approach to the microcanonical ensemble. In
contrast to the derivation of the canonical ensemble, the
maximization for the traditional approach to the microcanonical
ensemble is performed on a sub-manifold of the Hilbert space
(phase space in the classical case) rather than on the whole
Hilbert (phase) space. Thus the microcanonical ensemble is assumed
in the traditional approach rather than derived. In our approach
we have used the Hertz volume entropy with constraints of
normalization \emph{and} average energy to show that it leads to
the correct microcanonical distribution.

Some of the significant differences between the traditional
approach to the MEP in Eq. (\ref{eq:fakeMEP}) and our approach to
the MEP in Eqs. (\ref{Sconstr}) and (\ref{SConstr}) for the
microcanonical ensemble are the following.

\begin{enumerate}
\item The traditional MEP \emph{assumes} that the microstate $i$
belongs to energy eigenvalue $E_{\beta }$, whereas our MEP
\emph{derives} such condition. \item The traditional approach
employs the Shannon entropy \emph{without} an energy constraint,
whereas we employ the Hertz volume entropy \emph{with} the energy
constraint.

\item The traditional MEP derives Laplace's Principle of
Insufficient Reason for the states belonging to the eigenenergy
$E_{\beta }$, whereas our MEP invokes Laplace's principle after
deriving the condition that the state $i$ must belong to $E_{\beta
}$.
\end{enumerate}

In the latter case we see that our approach, rather than being in
contrast with the traditional one, \emph{completes} it. First one
maximizes the Hertz entropy to select the microcanonical energy
level. At this point can use the traditional method to find that
all states with that energy have the same probability.


\begin{thebibliography}{99}
\bibitem{jaynes1} E. T. Jaynes, Phys. Rev. 106 (1957) 620--630.

\bibitem{jaynes2} E. T. Jaynes, Phys. Rev. 108 (1957) 171--190.

\bibitem{tsallis1} C. Tsallis, J. Stat. Phys. 52 (1988) 479.

\bibitem{hertz} P. Hertz, \emph{Uber die mechanischen Grundlagen der
Thermodynamik}. Annalen der Physik (Leipzig) 33 (1910) 225-274 and 537-552.

\bibitem{gibbs} J. W. Gibbs, Elementary Principles in Statistical Mechanics,
New Haven, Yale University Press, 1902, p.170. Reprinted by Dover
Publications, New York, 1960.

\bibitem{boltz1} L. Boltzmann, Crelle's Journal 98 (1884) 68--94. Reprinted
in Hasen\"{o}hrl (ed.), Wissenschaftlic Abhandlungen, vol. 3. New York,
Chelsea, pp. 122-152.

\bibitem{campisi1} M. Campisi, Studies in History and Philosphy of Modern
Physics 36 (2005) 275--290.

\bibitem{gallavotti} G. Gallavotti, Statistical mechanics. A short treatise,
Springer Verlag, Berlin, 1995.

\bibitem{Ruelle} D. Ruelle, Statistical mechanics: rigorous results, New
York, W. A. Benjamin, 1969.

\bibitem{campisi-dual} M.Campisi, Physica A 385 (2007) 501-517,

\bibitem{campisibagci} M. Campisi and G. B. Bagci, Phys. Lett. A 362 (2007)
11-15.

\bibitem{campisiPLA2} M. Campisi, Phys. Lett. A, 366 (4-5) (2007) 335-338.





\bibitem{khin} A. I. Khinchin, Mathematical foundations of statistical
mechanics, Dover Publications, New York, 1949.

\bibitem{gurarie} V. Gurarie, Am. J. Phys. 75 (2007) 747-751.
\end{thebibliography}
\end{document}